\documentstyle{article}
\newread\epsffilein    
\newif\ifepsffileok    
\newif\ifepsfbbfound   
\newif\ifepsfverbose   
\newdimen\epsfxsize    
\newdimen\epsfysize    
\newdimen\epsftsize    
\newdimen\epsfrsize    
\newdimen\epsftmp      
\newdimen\pspoints     
\def\epsfbox#1{\global\def\epsfllx{72}\global\def\epsflly{72}%
   \global\def\epsfurx{540}\global\def\epsfury{720}%
   \def\lbracket{[}\def\testit{#1}\ifx\testit\lbracket
   \let\next=\epsfgetlitbb\else\let\next=\epsfnormal\fi\next{#1}}%
\def\epsfgetlitbb#1#2 #3 #4 #5]#6{\epsfgrab #2 #3 #4 #5 .\\%
   \epsfsetgraph{#6}}%
\def\epsfnormal#1{\epsfgetbb{#1}\epsfsetgraph{#1}}%
\def\epsfgetbb#1{%
\openin\epsffilein=#1
\ifeof\epsffilein\errmessage{I couldn't open #1, will ignore it}\else
   {\epsffileoktrue \chardef\other=12
    \def\do##1{\catcode`##1=\other}\dospecials \catcode`\ =10
    \loop
       \read\epsffilein to \epsffileline
       \ifeof\epsffilein\epsffileokfalse\else
          \expandafter\epsfaux\epsffileline:. \\%
       \fi
   \ifepsffileok\repeat
   \ifepsfbbfound\else
    \ifepsfverbose\message{No bounding box comment in #1; using defaults}\fi\fi
   }\closein\epsffilein\fi}%
\def\epsfsetgraph#1{%
   \epsfrsize=\epsfury\pspoints
   \advance\epsfrsize by-\epsflly\pspoints
   \epsftsize=\epsfurx\pspoints
   \advance\epsftsize by-\epsfllx\pspoints
   \epsfsize\epsftsize\epsfrsize
   \ifnum\epsfxsize=0 \ifnum\epsfysize=0
      \epsfxsize=\epsftsize \epsfysize=\epsfrsize
     \else\epsftmp=\epsftsize \divide\epsftmp\epsfrsize
       \epsfxsize=\epsfysize \multiply\epsfxsize\epsftmp
       \multiply\epsftmp\epsfrsize \advance\epsftsize-\epsftmp
       \epsftmp=\epsfysize
       \loop \advance\epsftsize\epsftsize \divide\epsftmp 2
       \ifnum\epsftmp>0
          \ifnum\epsftsize<\epsfrsize\else
             \advance\epsftsize-\epsfrsize \advance\epsfxsize\epsftmp \fi
       \repeat
     \fi
   \else\epsftmp=\epsfrsize \divide\epsftmp\epsftsize
     \epsfysize=\epsfxsize \multiply\epsfysize\epsftmp   
     \multiply\epsftmp\epsftsize \advance\epsfrsize-\epsftmp
     \epsftmp=\epsfxsize
     \loop \advance\epsfrsize\epsfrsize \divide\epsftmp 2
     \ifnum\epsftmp>0
        \ifnum\epsfrsize<\epsftsize\else
           \advance\epsfrsize-\epsftsize \advance\epsfysize\epsftmp \fi
     \repeat     
   \fi
   \ifepsfverbose\message{#1: width=\the\epsfxsize, height=\the\epsfysize}\fi
   \epsftmp=10\epsfxsize \divide\epsftmp\pspoints
   \vbox to\epsfysize{\vfil\hbox to\epsfxsize{%
      \special{illustration #1 scaled \number\epsfscale}
      \hfil}}%
\epsfxsize=0pt\epsfysize=0pt\epsfscale=1000 }%

{\catcode`\%=12 \global\let\epsfpercent=
\long\def\epsfaux#1#2:#3\\{\ifx#1\epsfpercent
   \def\testit{#2}\ifx\testit\epsfbblit
      \epsfgrab #3 . . . \\%
      \epsffileokfalse
      \global\epsfbbfoundtrue
   \fi\else\ifx#1\par\else\epsffileokfalse\fi\fi}%
\def\epsfgrab #1 #2 #3 #4 #5\\{%
   \global\def\epsfllx{#1}\ifx\epsfllx\empty
      \epsfgrab #2 #3 #4 #5 .\\\else
   \global\def\epsflly{#2}%
   \global\def\epsfurx{#3}\global\def\epsfury{#4}\fi}%
\newcount\epsfscale    
\newdimen\epsftmpp     
\newdimen\epsftmppp    
\newdimen\epsfM        
\newdimen\sppoints     
\epsfscale=1000        
\sppoints=1000sp       
\epsfM=1000\sppoints
\def\computescale#1#2{%
  \epsftmpp=#1 \epsftmppp=#2
  \epsftmp=\epsftmpp \divide\epsftmp\epsftmppp  
  \epsfscale=\epsfM \multiply\epsfscale\epsftmp 
  \multiply\epsftmp\epsftmppp                   
  \advance\epsftmpp-\epsftmp                    
  \epsftmp=\epsfM                               
  \loop \advance\epsftmpp\epsftmpp              
    \divide\epsftmp 2                           
    \ifnum\epsftmp>0
      \ifnum\epsftmpp<\epsftmppp\else           
        \advance\epsftmpp-\epsftmppp            
        \advance\epsfscale\epsftmp \fi          
  \repeat
  \divide\epsfscale\sppoints}
\def\epsfsize#1#2{%
  \ifnum\epsfscale=1000
    \ifnum\epsfxsize=0
      \ifnum\epsfysize=0
      \else \computescale{\epsfysize}{#2}
      \fi
    \else \computescale{\epsfxsize}{#1}
    \fi
  \else
    \epsfxsize=#1
    \divide\epsfxsize by 1000 \multiply\epsfxsize by \epsfscale
  \fi}

\def\idem{\smallskip\noindent
\hangindent 2 pc }

\font\tenbg=cmmib10 at 10pt

\def \rvecphi{{\hbox{\tenbg\char'036}}}

\begin{document}

\title{Relativistic Jets from
Accretion Disks}

\author{R.V.E. Lovelace, P.R. Gandhi, \& M.M. Romanova\\
Cornell University}

\maketitle

\begin{abstract} 

  The jets observed to emanate from many 
compact accreting objects may arise from the
twisting of a magnetic field threading a 
differentially rotating accretion disk which 
acts to magnetically extract angular
momentum and energy from the disk. 
  Two main regimes have been discussed, hydromagnetic
jets, which have a significant mass flux 
and have energy and angular momentum carried
by both matter and electromagnetic field and, 
Poynting jets, where the mass flux is
small and energy and angular momentum 
are carried predominantly by the
electromagnetic field. 
  Here, we describe recent
theoretical work on the formation of
relativistic Poynting jets from magnetized accretion disks.
   Further, we describe
new relativistic, fully-electromagnetic,
particle-in-cell simulations  of the formation
of  jets from accretion disks. 
   Analog
Z-pinch experiments may help
to understand the origin of astrophysical jets.

\end{abstract}

\medskip

\centerline{\bf 1. INTRODUCTION}
\medskip
Powerful, highly-collimated, oppositely 
directed jets are observed in active galaxies 
and quasars (see for example Bridle \& Eilek
1984), and in old compact stars in binaries - the
``microquasars'' (Mirabel \& Rodriguez 1994; 
Eikenberry {\it et al.} 1998). Further, highly
collimated emission line jets are seen 
in young stellar objects (B\"uhrke, Mundt, \& Ray
1988). 
   Different models have been put 
forward to explain astrophysical jets
(Bisnovatyi-Kogan \& Lovelace 2001). 
   Recent observational and theoretical work favors
models where twisting of an ordered 
magnetic field threading an accretion disk acts to
magnetically accelerate the jets. 
   Here, we discuss the origin of the relativistic jets
observed in active galaxies and quasars
and in microquasars.  We first discuss a theoretical
model (\S 1), and then new results from relativistic
particle-in-cell (PIC) simulations (\S 2).

\medskip

\centerline{\bf 2. POYNTING JETS}
\medskip

The powerful jets observed from active 
galaxies and quasars are probably not
hydromagnetic outflows but rather Poynting flux
dominated jets. 
  The motions of these jets measured
by very long baseline interferometry correspond
to bulk Lorentz factors of $\Gamma = {\cal O}(10)$ 
which is much larger than the Lorentz
factor of the Keplerian disk velocity
predicted for hydromagnetic outflows.
  Furthermore, the low Faraday rotation measures
observed for these jets at distances $<$ kpc from
the central object implies a very low plasma
densities. Similar arguments indicate that the
jets of microquasars are not hydromagnetic
outflows but rather Poynting jets. Poynting jets
have also been proposed to be the driving mechanism for
gamma ray burst sources (Katz 1997). 
Theoretical studies have developed models for
Poynting jets from accretion disks (Lovelace,
Wang, \& Sulkanen 1987; Lynden-Bell 1996;
Romanova \& Lovelace 1997; Levinson 1998; 
Li {\it et al.} 2001; 
Lovelace {\it et al.} 2002; and Lovelace
\& Romanova 2003). 
   Stationary non-relativistic Poynting flux
dominated outflows were found by Romanova {\it et al.}
(1998) and Ustyugova {\it et al.} (2000) in
axisymmetric MHD simulations of the opening of
magnetic loops threading a Keplerian disk.
Here, we summmarize a model for the formation
of relativistic Poynting jets from a disk
(Lovelace \& Romanova 2003).

		Consider a dipole-like coronal magnetic 
field - such as that shown in the lower part of
Figure 1a - threading a differentially rotating Keplerian
accretion disk. 
  The disk is perfectly
conducting, high-density, and has a small
accretion speed ($\ll$ Keplerian speed). 
   The field may be generated in the disk by
a dynamo action and released.
    Outside of the disk there is assumed
to be a 
``coronal'' or ``force-free" plasma 
($\rho_e{\bf E}+{\bf J \times B}/c\approx 0$, Gold \& Hoyle 1960). 
  We use cylindrical $(r, \phi,
z)$ coordinates and consider axisymmetric field
configurations. Thus the magnetic field has the
form ${\bf B} = {\bf B}_p + B_\phi\hat{\rvecphi}$, 
with ${\bf B}_p =
B_r\hat{\bf r} + B_z\hat{\bf z}$. 
We have $B_r = -(1/r)\partial \Psi/\partial z$
and  $B_z = (1/r)\partial \Psi/\partial r$.
where $\Psi(r,z) \equiv rA_\phi(r,z)$ is the flux
function.

    Most of the azimuthal twist $\Delta \phi$ of a field line
of the Poynting  jet
	occurs along the jet from $z = 0$ to $Z(t)$ as sketched in
Figure 1a, where $Z(t)$ is the
	axial location of the ``head'' of the jet. 
   Along most of the
distance $z = 0$ to $Z$, the radius 
of the jet is a constant and 
$\Psi = \Psi(r)$ for $Z >> r_0$, where $r_0$ 
is the radius of the O-point of the magnetic 
field in the disk.
   Note that the function $\Psi(r)$
is different from $\Psi(r,0)$ which 
is the flux function profile
on the disk surface. 
Hence $r^2 d\phi/dz =rB_\phi(r,z)/B_z(r,z)$.
 We take for simplicity  $V_z =
dZ/dt =$ const. We determine $V_z$ subsequently. 
  In this case  $H (\Psi) =[r^2\Omega(\Psi)/V_z]B_z$
can be written as a function of $\Psi$ and
$d\Psi/dr$.  With $H$ known, the relativistic 
Grad-Shafranov equation,
$$
\left[1-\left({r\Omega \over c}\right)^2\right]
\Delta^*\Psi -{{\bf \nabla \Psi}\over 2r^2}\cdot
{\bf \nabla}\left({r^4\Omega^2 \over c^2}\right)=
-H(\Psi){d H(\Psi) \over d \Psi}~,
\eqno(1)
$$
can be solved (Lovelace \& Romanova 2003).

	  The quantity not determined by 
equation (1) is the velocity $V_z$, or Lorentz factor 
$\Gamma =1/(1-V_z^2/c^2)^{1/2}$.
 This is determined by taking into account the balance of
axial forces at the head of the jet: the electromagnetic 
pressure within the jet is balanced against the
dynamic pressure of the external medium 
which is assumed uniform with density $\rho_{ext}$.
This gives $(\Gamma^2-1)^3=B_0^2/
(8\pi {\cal R}^2\rho_{ext} c^2)$, or for $\Gamma \gg 1$,
$$
\Gamma \approx 8\left({10 \over {\cal R}}\right)^{1/3}
\left({B_0 \over 10^3 {\rm G}}\right)^{1/3}
\left({1/{\rm cm}^3 \over n_{ext}}\right)^{1/6}~,
\eqno(2)
$$
where ${\cal R} = r_0/r_g \gg 1$ and
 $r_g \equiv 	GM/c^2$, and
$B_0$ the magnetic field strength at the 
center of the disk.   A necessary condition for the validity of
this equation is that the axial speed
of the counter-propagating
 fast magnetosonic wave (in the lab
frame) be larger than
$V_z$ so that the jet is effectively
`subsonic.'
  This value of $\Gamma$ is of the
order of the Lorentz factors of the expansion of
parsec-scale extragalactic radio jets observed
with very-long-baseline-interferometry 
(see, e.g., Zensus {\it et al.} 1998). 
This interpretation assumes that the
radiating electrons (and/or positrons) 
are accelerated to high Lorentz
factors ($\gamma \sim 10^3$) at the 
jet front and move with a
bulk Lorentz factor $\Gamma$ relative to the
observer. 
   The luminosity of the $+z$ Poynting 
jet is $\dot{E}_j =c\int_0^{r_0} 
rdrE_rB_\phi/2= cB_0^2{\cal R}^{3/2}r_g^2/3\sim
2.1 \times 10^{46} (B_0/ 10^3{\rm G})^2 
({\cal R}/10)^{3/2}(M/10^9M_\odot)^2$ erg/s, 
where $M$ is the mass of
the black hole.

 For long time-scales, the Poynting jet 
is time-dependent due to the 
angular momentum it extracts from
the inner disk ($r < r_0$) which in turn 
causes $r_0$ to decrease with time
(Lovelace {\it et al.} 2002). 
  This loss of angular momentum 
leads to a ``global
magnetic instability'' and collapse 
of the inner disk (Lovelace {\it et al.}
1994, 1997, 2002) and a corresponding 
outburst of energy in the jets from
the two sides of the disk. 
  Such outbursts may explain the flares of
active galactic nuclei blazar sources 
(Romanova \& Lovelace 1997; Levinson
1998) and the one-time outbursts of 
gamma ray burst sources (Katz 1997).

\medskip
\centerline{\bf 3. RELATIVISTIC ELECTROMAGNETIC PIC SIMULATIONS}
\medskip

    We performed relativistic, fully 
electromagnetic, particle-in-cell
simulations of the formation of jets from  an accretion disk
initially threaded by a dipole-like magnetic field.
   This was done using the code XOOPIC developed
by Verboncoeur, Langdon, and Gladd (1995).
  Earlier, Gisler, Lovelace, and Norman (1989)
studied jet formation for a monopole type field using
the relativistic PIC code ISIS. 
   The geometry of the initial configuration is 
shown in Figure 1b.
    The computational region is a cylindrical ``can,''
$r=0 -R_m$ and $z=0 - Z_m$, with outflow boundary conditions
on the outer boundaries, and the potential and particle
emission specified on the disk surface $r=0-R_m$, $z=0$. 
  Equal fluxes of electrons and positrons are emitted
so that the net emission  is 
effectively  space-charge-limited.
  About $10^5$ particles were used in the simulations
reported here.
   The behavior of the lower half-space ($z<0$) is expected to
be a mirror image of the upper half-space.

   Figure 2 shows the formation
of a relativistic jet. 
The gray scale indicates the logarithm
of the density of electrons or positrons with $20$ levels
between the lightest ($10^{12}$) 
and darkest ($4\times 10^{15}$/m$^3$).  
   The lines are poloidal
magnetic field lines ${\bf B}_p$.  The total
${\bf B}-$field is  shown in Figure 3.
  The computational region has
$(R_m,Z_m)=(50,100)$ m, the initial ${\bf B}-$field is
dipole-like  with
$B_z(0,0)\equiv B_0=28.3$ G and an O-point at
$(r,z)=(10, 0)$ m, and the electric
 potential at the center of the disk 
is $\Phi_0=-10^7$ V relative to the outer region of the
disk.  Initially, the computational region was filled
with a distribution of equal densities of electrons
and positrons with $n_\pm(0,0) =3\times 10^{13}$/m$^3$.
  Electrons and positrons
are emitted with equal currents $I_{\pm} = 3\times 10^5$ A
from both the inner and the outer 
portions of the disk as indicated in Figure 1b with an
axial speed much less than $c$. 
   For a Keplerian disk with $r_0 \gg r_g$, the scalings are $\Phi_0
\sim  B_0 (r_0 r_g)^{1/2}$, 
$I\sim cB_0 r_0$ and the jet power is $\sim cB_0^2 
r_0^{3/2}r_g^{1/2}$. The calculations 
were done on a $64\times 128$ grid stretched in
both the $r$ and $z$ directions so as to give
much higher spatial resolution at small $r$ and small $z$.
  These simulations show the formation of a quasi-stationary,
collimated current-carrying jet.   The Poynting flux
power of the jet is $\dot{E}_j\approx 7\times 10^{11}$ W and the
particle kinetic energy power is $\approx 4.7 \times 10^{10}$ W.
The charge density
of the electron flow is partially neutralized by
the positron flow.
    Simlations are planned with the positrons
replaced by ions.
  Scaled Z-pinch experiments configured as shown in Figure 1b can
allow further study of  astrophysical jet formation.

  We thank C. Birdsall, S. Colgate,
 H. Li, J. Verboncoeur, I. Wasserman,
J. Wick, and T. Womack 
for valuable assistance and discussions.
  This work was supported in part 
by NASA grants NAG5-13060
and NAG5-13220, by NSF grant AST-0307817, 
and by DOE cooperative agreement DE-FC03
02NA00057.

\smallskip
\noindent{\bf REFERENCES}
\smallskip

\idem Bisnovatyi-Kogan, G.S. \& Lovelace, R.V.E. 2001, New
Astron. Rev., 45, 663 

\idem Bridle, A.H., \& Eilek, J.A.
(eds.) 1984, in Physics of Energy Transport in
Extragalactic Radio Sources, (Greenbank:- NRAO)

\idem B\"uhrke, T., Mundt, R., \& Ray, T.P. 1988, A\&A, 200, 99

\idem Eikenberry, S., Matthews, K., Morgan, E.H., 
Remillard, R.A., \& Nelson, R.W. 1998,
ApJ, 494, L61

\idem Gisler,ÊG., Lovelace, R.V.E., \& Norman,ÊM.L. 1989,
ApJ, 342, 135

\idem Gold, T., \& Hoyle, F. 1960, MNRAS, 120, 89

\idem Katz, J.1. 1997, ApJ, 490, 633

\idem Levinson, A. 1998, ApJ, S07, 145

\idem Li, H., Lovelace, R.V.E., Finn, J.M., 
\& Colgate, S.A. 2001, ApJ, 561, 915

\idem Lovelace, R.V.E., Li, H., Koldoba, A.V, 
Ustyugova, G.V, \& Romanova, M.M. 2002,
ApJ,572,445

\idem Lovelace, R.V.E., Newman, W.I., 
\& Romanova, M.M. 1997, ApJ, 484,
628

\idem Lovelace, R.V.E., Romanova, M.M., 
\& Newman, W.I. 1994, ApJ, 437,
136

\idem Lovelace, R.V.E., Wang, J.C.L., 
\& Sulkanen, M.E. 1987, ApJ, 315,
504

\idem Lovelace, R.V.E., \& Romanova, M.M. 2003, 
ApJ, 596, L159

\idem Lynden-Bell, D. 1996, MNRAS, 279, 389

\idem Mirabel, I.F., \& Rodriguez, L.F. 1994 Nature, 371, 46

\idem Romanova M.M., \& Lovelace R.V.E. 1997, ApJ, 475, 97

\idem Romanova, M.M., Ustyugova, G.V, Koldoba, A.V, 
Chechetkin, VM., \& Lovelace,
R.V.E. 1998, ApJ, 500, 703

\idem Ustyugova, G.V, Lovelace, R.V.E., Romanova, M.M., 
Li, H., \& Colgate, S.A. 2000
ApJ, 541, L21

\idem Verboncoeur, J.P., Langdon, A.B., \& Gladd, N.T. 1995,
Comp. Phys. Comm., 87, 199

\idem Zensus, J.A., Taylor, G.B., \& Wrobel, J.M. (eds.) 1998, 
Radio Emission from Galactic and Extragalactic 
Compact Sources, IAU Colloquium
164, (Ast. Soc. of the Pacific)

\begin{figure*}[t]
\epsfysize=6.5cm 
\centerline{\epsfbox{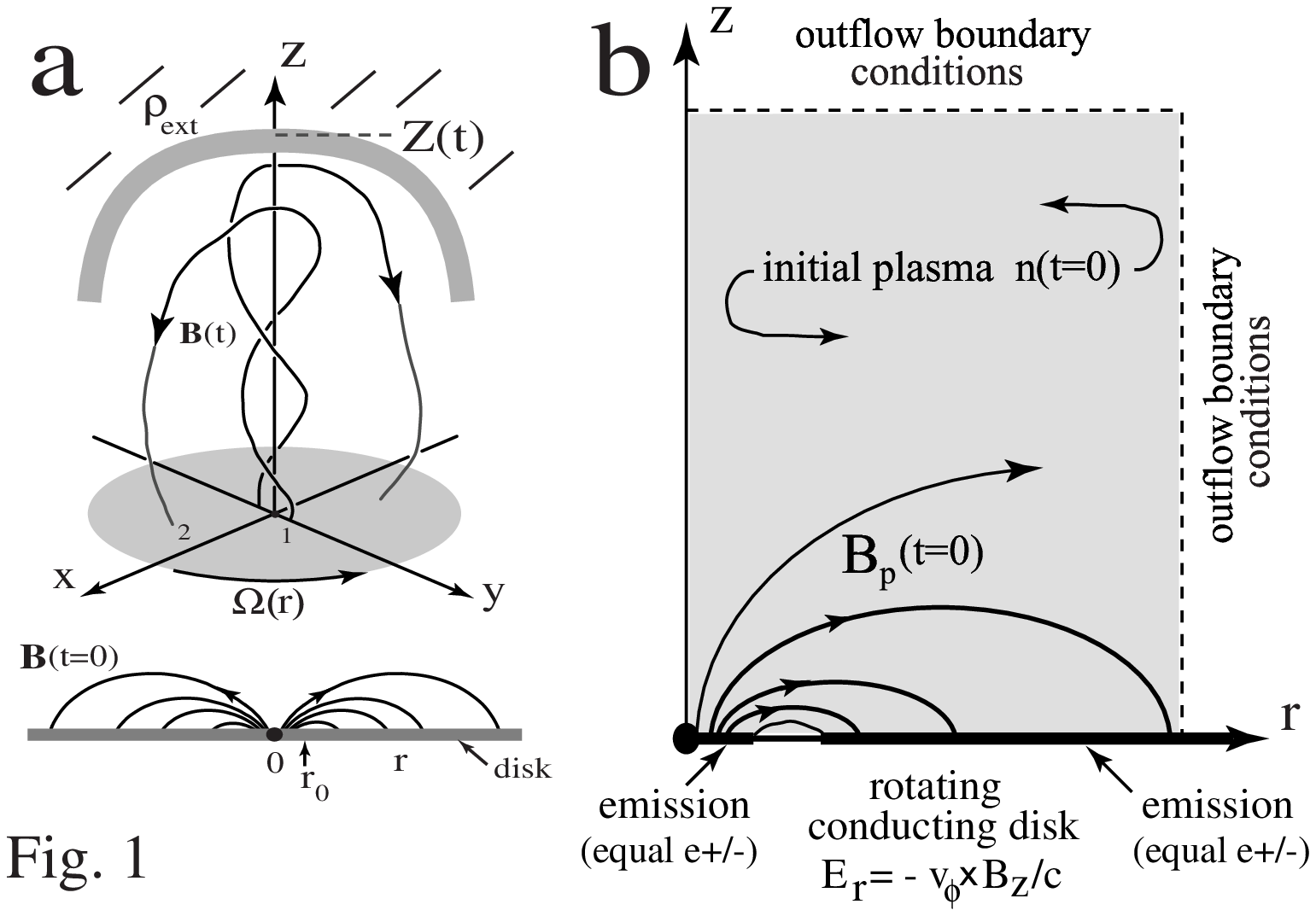}}
\caption{
({\bf a}) Sketch of the magnetic field 
configuration of a Poynting jet from Lovelace and
Romanova (2003). The bottom part of ({\bf a})
shows the initial dipole-like magnetic field
threading the disk which rotates at the angular
rate $\Omega(r)$. The top part  shows
the jet at some time later when the head of the
jet is at a distance $Z(t)$. At the head of the jet
there is force balance between electromagnetic
stress of the jet and the ram pressure of the
ambient medium of density $\rho_{ext}$.
({\bf b}) Sketch of the initial conditions
for the relativistic PIC simulations of 
jet formation from an accretion disk.}
\end{figure*}

\begin{figure*}[t]
\epsfysize=11.cm 
\centerline{\epsfbox{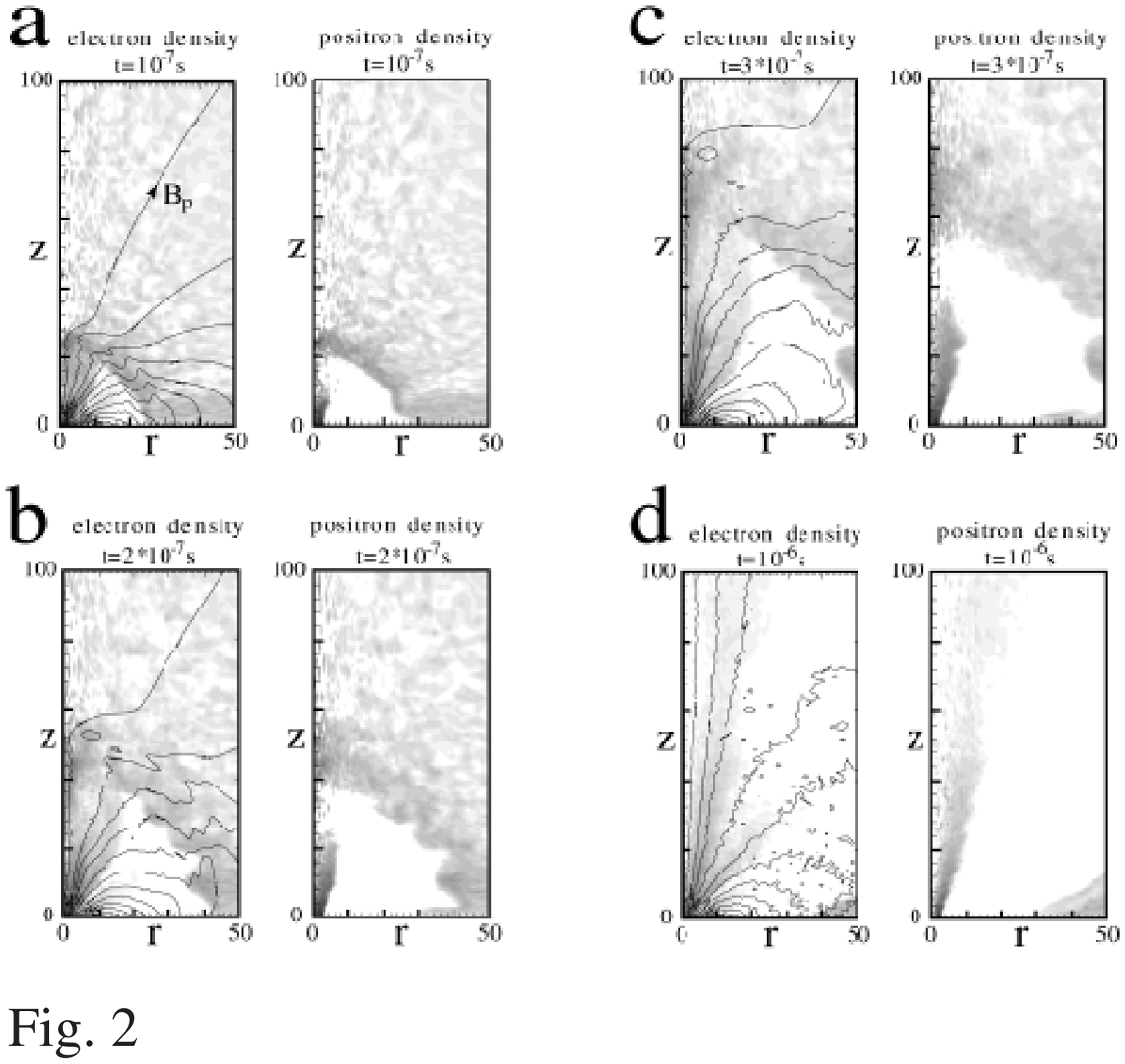}}
\caption{Relativistic PIC simulations of
the formation of a jet from a rotating disk.
(a) -(c) give snapshots at times $(1,2,3)\times10^{-7}$ s,
and (d) is at $t=10^{-6}$ s.
}
\end{figure*}

\begin{figure*}[t]
\epsfysize=12cm 
\centerline{\epsfbox{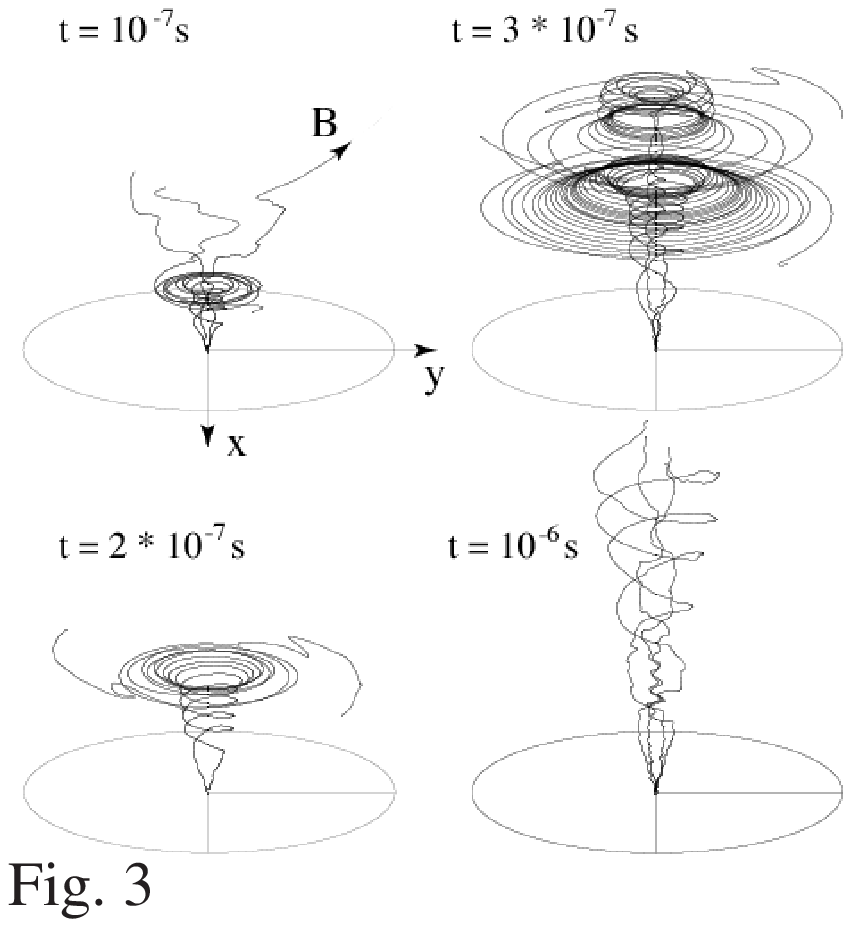}}
\caption{Three dimensional magnetic
field lines originating from the disk
at $r=1, 2$ m  for the same case as Figure 2.}
\end{figure*}

\end{document}